# Evaluating Optical Fiber Links with Data Filtering and Allan Deviation

Claudio E. Calosso, Cecilia Clivati, Salvatore Micalizio
Physics Metrology Department
Istituto Nazionale di Ricerca Metrologica, INRIM
Torino, Italy
e-mail: c.calosso@inrim.it

*Abstract* — In this paper we propose a simple method to reject the high-frequency noise in the evaluation of statistical uncertainty of coherent optical fiber links. Specifically, we propose a preliminary data filtering, separated from the frequency stability computation. In this way, it is possible to use the Allan deviation as estimator of stability, to get unbiased data, which are representative of the noise process affecting the delivered signal. Our approach is alternative to the use of the modified Allan deviation, which is largely adopted in this field. We apply this processing to the experimental data we obtained on a 1284 km coherent optical link for frequency dissemination, which we realized in Italy. We also show how the so-called Lambda-type commercial phase/frequency counters can be used to this purpose.

## I. INTRODUCTION

In recent years, coherent optical fiber links for frequency dissemination have become an established practice [1-5] and a growing number of fiber-based atomic clocks comparisons are going to be performed in next years [6-8]. This is motivated by the increased accuracy and stability of this technique as compared to state-of-the-art satellite links: optical links achieve a resolution at the level of $1\times10^{-19}$ after one day of averaging and are likely to replace satellite techniques at least on continental scales. This opens new possibilities not only in metrology, but also in fundamental physics [9], high-precision spectroscopy [10], geodesy [11] and radioastronomy [12].

The coherent frequency transfer via optical fiber is based on the delivery of an ultrastable laser at telecom wavelength along a standard fiber for telecommunications, where the length variations are actively canceled through the Doppler-noise-cancellation scheme [13] or rejected by means of the 2-way technique [14]. With these schemes, the intrinsic noise of thousands-kilometers long fibers, due to vibrations and temperature changes, can be reduced up to a bandwidth of tens of hertz. The residual phase noise of the delivered laser actually depends on the intrinsic noise of the fiber itself and on the fiber length [15]. The several experimental realizations so far implemented all have some common features, namely a compensation bandwidth $B_L$ of few tens of hertz, determined by the link length $L$ through the relation $B_L = c_n/(4L)$ ($c_n$ being the speed of light in the fiber), and a phase noise of type $S_\varphi(f) \propto f^0$ (white phase noise, WPN) or $S_\varphi(f) \propto f^1$ (blue phase noise BPN) [2,3] (here, we assume that the phase noise can be described by the power law function $S_\varphi(f) = \sum_\alpha b_\alpha f^\alpha$, $-4 < \alpha \leq +1$). At very low Fourier frequencies ($f < 0.01$ Hz), the uncompensated fiber noise is negligible as compared to some long-term effects on the interferometer and to the intrinsic noise of the clocks to be compared. Nevertheless, the residual fiber noise always affects fiber-based frequency dissemination and remote clocks comparisons, as the fiber noise typically extends up to few kilohertz of Fourier frequency. In addition, a strong bump has been reported in many realizations [1-4] between 10 Hz and 30 Hz. This is due to acoustic noise and vibrations on the optical fiber, which are not completely compensated by the noise-cancellation scheme at these frequencies.

Of course, one is interested in rejecting such high-frequency noise, since, in general, it does not contain useful information. Although these noise components can be effectively discriminated in a frequency-domain measurement, it is not straightforward to identify and reject them in a time-domain measurement. The resulting effect is well know when measuring the performances of optical links with frequency counters. If a proper procedure is not adopted, the traditional estimator for statistical uncertainty, the Allan deviation (ADEV) [16], is saturated by high-frequency noise and, as a consequence, the stability is dominated by the link rather than by the intrinsic noise of the clocks.

In this work, we describe how a proper use of the ADEV can avoid this effect. Namely, we show how filtering the original data effectively rejects the high-frequency noise of the link, while it does not affect the ADEV behavior in the long term. This approach has the advantage of optimizing the ADEV computation to extract the maximum information from experimental data. In addition, it allows a direct and unambiguous comparison of the results, being in continuity with the existing literature, and it is capable, at the same time, to cope with the issues posed by the new frequency comparison techniques.

We note that it has become a common practice to use the Modified Allan deviation (MDEV) [17] as an estimator for frequency stability in optical links [2,4,8]: the MDEV has the advantage of mitigating the effect of the link high-frequency noise already at short averaging time. As a drawback, the estimation of the clocks stability (which we assume dominated by white-frequency noise, WFN) is affected as well and the ultimate stability no longer corresponds to the classical deviation [17,18]. We stress that this correspondence is the underlying reason behind the choice of the ADEV as a stability estimator in the metrological community [19]. Thus, the choice of the MDEV has to be motivated, and recent works are addressing this issue [20].

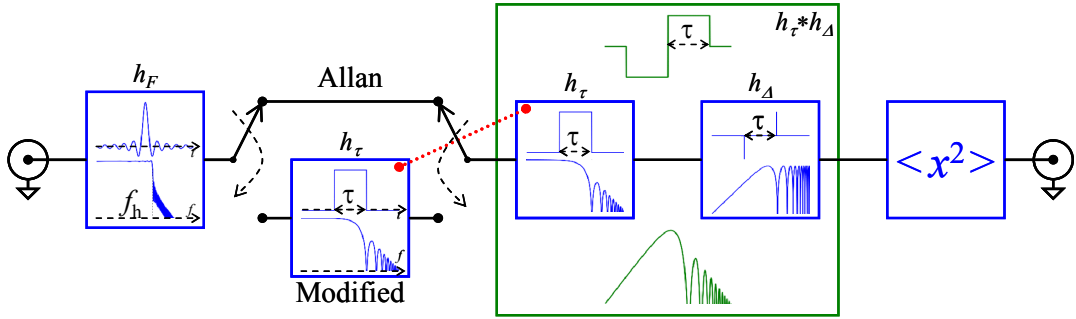

Figure 1: the block diagrams of the ADEV and of the MDEV. The first block $h_F$ represents a filtering process with bandwidth $f_h$ (both impulse response and transfer function are shown); the switch highlights the presence of the additional moving average filter $h_\tau$ in the MDEV, whose bandwidth changes with the averaging time $\tau$; the green block represents the well-known ADEV equivalent filter, i.e. the average $h_\tau$ over a time $\tau$ and the difference $h_\Delta$ between adjacent samples; the last block represents evaluation of the mean power.

In the following sections, we will describe our approach; in particular, we will separate the counting process from the ADEV computation and propose an experimental procedure to get unbiased data, truly representative of the experimental noise process. Although phasemeters are the best instruments to perform such measurements [21,22], we will show how the most common commercial phase and frequency counting devices can be easily adapted to this task [23]. We then propose an experimental case where this method can be applied, showing the results we obtained on the 1284 km optical fiber link for frequency dissemination developed by the National Metrology Institute in Italy (INRIM) [3].

## II. THEORY

We suppose we have a working optical frequency link and we want to characterize its performance; this is routinely made by comparing two links with the same starting and ending points [4] or by looping a single link, so that the remote end coincides with the local end [1-3]. The output of the measurement is a stream of dead-time free phase or frequency data that are supposed to be processed by statistical tools.

Every measurement system has a finite bandwidth that can be modeled by a low-pass filter with an equivalent bandwidth $f_h$. Ideally, this filter will completely reject the noise above $f_h$ and can be either hardware or software realized. Hereon, we refer to this operation with the term pre-filtering.

The data are then processed by a statistical estimator. Specifically, we focus onto the two typical signal analysis methods commonly adopted in frequency metrology, the ADEV and the MDEV.

The ADEV is a well known estimator and is universally accepted in frequency metrology. It is described by the formula

$$\sigma_y(\tau) = \sqrt{\left\langle \frac{1}{2}(\bar{y}(t) - \bar{y}(t-\tau))^2 \right\rangle}$$
$$= \sqrt{\left\langle \frac{1}{2}\left(\frac{dx(t)}{dt} * h_F(t) * h_\tau(t) * h_\Delta(t)\right)^2 \right\rangle}$$

where $\bar{y}(t)$ represents the average relative frequency over the time interval [t−τ, t] and τ is the averaging time, the brackets $\langle \ \rangle$ denote an infinite time average, and $h_F$, $h_\tau$, $h_\Delta$ are the impulse response of the pre-filter that sets the measurement bandwidth, of the average over τ and of the difference respectively. As usual, $y = \frac{dx}{dt}$ where $x$ is the phase time [14-15].

Historically, the MDEV has been introduced to deal with fast noise processes, such as WPN. It is defined by the formula

$$\text{Mod } \sigma_y(\tau) = \sqrt{\left\langle \frac{1}{2}\left(\frac{1}{n}\sum_{i=1}^{n}\bar{y}(t-i\tau_0) - \frac{1}{n}\sum_{i=1}^{n}\bar{y}(t-\tau-i\tau_0)\right)^2 \right\rangle}$$
$$\cong \sqrt{\left\langle \frac{1}{2}\left(\frac{dx(t)}{dt} * h_F(t) * h_\tau(t) * h_\tau(t) * h_\Delta(t)\right)^2 \right\rangle}$$

It can be seen that, for $\tau \gg \tau_0$, the MDEV is equivalent to the ADEV with the introduction of the additional moving-average filter $h_\tau$, as the one used in ADEV. Here, $\tau_0$ represents the shortest averaging time, i.e. the original gate time of the acquisition device. The block diagram of the two estimators is sketched in Fig. 1, where the transfer functions of ADEV is highlighted as well.

We stress the importance of the preliminary filtering block, which is separated from the uncertainty evaluation. Filtering data allows the rejection of the high frequency noise which would otherwise saturate the ADEV; then, the ADEV can be applied straightforwardly on filtered data, without the need of other estimators.

Filtering is applied once at the beginning of the data analysis and does not affect the measurement on long averaging times, where, in many cases, the noise is dominated by processes with $\alpha \leq -1$. In addition, we note that for $\alpha \leq -2$, the Allan deviation does not even depend on $f_h$: $f_h$ only establishes how much of the undesired link noise is rejected. A good choice of $f_h$ is thus the Fourier frequency where the measurement noise starts to be dominated by processes with $\alpha \leq -1$.

The filter selectivity can be tailored on the specific noise process encountered in the experiment. Again, this does not affect the stability on the long term. A higher selectivity might require a longer impulse response, however this is not an issue in most practical cases, as typically $f_h \gg 1/T$, where $T$ is the duration of the experiment.

We note that the additional filter in the MDEV definition has the same target of the pre-filter, i. e. to reject high-frequency noise. However, whilst the pre-filter has a fixed bandwidth $f_h$, the additional moving average changes its bandwidth $B$ with the measurement time, according to the relation $B = 1/(2\tau)$. This is the reason why the stability estimation is different for ADEV and MDEV. In fact, although this filter is introduced to reject noise on short measurement times, it acts at all timescales, in particular over long measurement times where, in most cases, it is not necessary. Then, the stability estimation turns out affected; specifically, the faster is the noise process, the higher is the bias. Table 1 and Fig. 2 summarize the bias for the various noise types, from WPN to frequency random walk and frequency linear drift. In Fig. 2, we show the behavior of the ADEV and MDEV (red and blue respectively). Two cases are considered: with (dashed lines) and without (continuous lines) white phase noise. In the inset, we show the noise that underlies the calculation of the deviations: it is composed of noise processes with $-4 \leq \alpha \leq 0$ and each one dominates for two decades.

| $\alpha$ | NOISE TYPE | $\overline{\text{MDEV}}/\text{ADEV}$ | dB |
|---|---|---|---|
| – | linear frequency drift | 1 | 0 |
| –4 | random walk FM | 0.908 | –0.84 |
| –3 | flicker FM | 0.822 | –1.71 |
| –2 | white FM | 0.707 | –3.01 |
| –1 | flicker PM | 0.404* | –7.84* |

*in case $f_h \tau = 100$.

Tab. 1: MDEV/ADEV for $\alpha \leq -1$ and linear frequency drift. The same amount is expressed in dB and has a meaning of underestimation or bias.

It can be seen that MDEV is unbiased in the presence of a linear frequency drift, while the bias increases with $\alpha$, i. e.

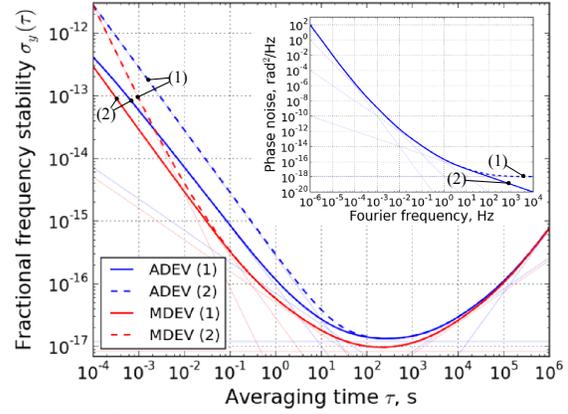

Fig. 2: ADEV (blue) and MDEV (red) provide different estimations, especially for fast noise processes. Here it is shown for the phase noise in the inset, with (dashed line) and without (continuous line) WPN.

with fast noise processes. In presence of WFN, which is the typical case of atomic clocks, the bias is −29% (−3 dB). Such bias is an issue especially in the latter case, because the uncertainty can no longer be related to the standard deviation as expected. Another consequence of the MDEV additional filter is that, in order to be applied over long measurement times, it must have a short impulse response, which implies a poor selectivity. The moving average is a compromise between selectivity at a given cutoff frequency and a sufficiently short impulse response.

The pre-filtering procedure which we propose enables to optimize the stability analysis so that both the requirement of efficient filtering and convergence are satisfied. In addition, it does not bias the long-term behavior of the estimator and it allows an unambiguous comparison of the results between different experiments and also with the literature, where the ADEV has been traditionally used. In fact, the ADEV has been analytically derived as a function of $f_h$ for the various noise types. Thus, if the noise process is known (this information can be retrieved by the corresponding spectra) it is straightforward to extrapolate the equivalent uncertainty for various $f_h$. In addition, this issue simply does not exist in the cases where $\alpha \leq -2$.

The procedure is then straightforward: the noise rejection is completely managed by the pre-filter, whose shape and $f_h$ are chosen by looking at the phase noise spectra. Then, the ADEV is applied to calculate the frequency stability.

We point out that the ADEV has no meaning for averaging times $\tau < 1/(2f_h)$ because of the filtering procedure. Fig. 3 shows the behavior of the ADEV in the case of white frequency noise. The ADEV is underestimated for $\tau f_h \ll 1$ and starts to be meaningful for $\tau = 1/(2f_h)$, where the underestimation is 20% (or 1.9 dB).

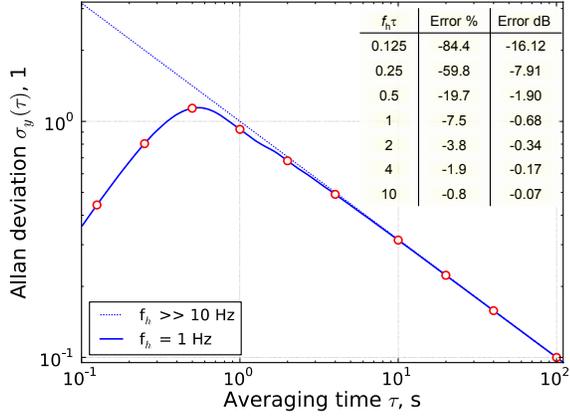

Fig. 3: the effect of filtering on the ADEV: the ADEV can be used for $\tau > 1/(2f_h)$. The underestimation with respect to an almost unfiltered case ($f_h \gg 10$ Hz) decreases quickly with $\tau$. The inset table shows the underestimation for various values of $f_h \tau$.

## III. COUNTING IN PRACTICE

It may be useful at this point to give a simple procedure for data acquisition in an experimental session. When quantifying the behavior of an optical link, the variable of interest is phase, as it is proportional to the optical length and it is directly subject to discrete cycle slips. It is then advisable to sample phase rather than frequency; this latter can always be derived from phase data unambiguously.

The first consideration is on the sampling rate, that must fulfill the Nyquist theorem [24]. For instance, thousands-kilometers long optical links are affected by phase noise up to 200 Hz or 500 Hz, hence the sampling rate should be around 1 kHz in order to avoid aliasing effects on the counting. Shorter links may have a servo-bump at higher frequency and the sampling frequency needs to be set accordingly, to avoid aliasing; a more detailed description the effect of aliasing is shown in Section IV.

The best instrument to perform this task is a phasemeter; however, also a more common and cheaper electronic counter can be used, provided that it is dead-time free or that it used in the time-interval mode, which is intrinsically dead-time free. We note however that a good knowledge of the counter operation is not trivial, as demonstrated by the broad literature [25,26] and care must be taken to avoid incurring in mistakes. From the theoretical point of view, both phase and frequency data contain the same information; the following analysis is based on phase data, as currently done in the characterization of signal distribution.

The second step is data filtering; the filter shape is not critical and several typologies have been described in the literature. The only requirements for the filters are: 1) a sufficient attenuation of the link noise; 2) a bandwidth $f_h$ set at a frequency where the clocks noise, rather than the link, starts to dominate the spectrum; 3) a unity gain at low frequency.

Although filtering could be analog-implemented, digital implementation offers a variety of possibilities and enables to simultaneously meet all requirements, tailoring the filter parameters on the specific experiment. The simplest filtering option is the moving average; although having a poor selectivity, it requires minimal computational effort, and it is already implemented in several commercial phase/frequency counters which operate in the so-called Lambda mode. However, when dealing with optical links, it may be useful to use sharper filters, to better reject the fiber noise. For instance, Fig. 4 compares the transfer function of a truncated sinc filter [24] to a moving average filter with the same equivalent bandwidth $f_h = 0.05\, f_s$, being $f_s$ the sampling frequency. At the Nyquist frequency they attenuate 65 dB and 20 dB respectively. A wide variety of digital filter algorithms are available in the most common libraries for digital signal processing.

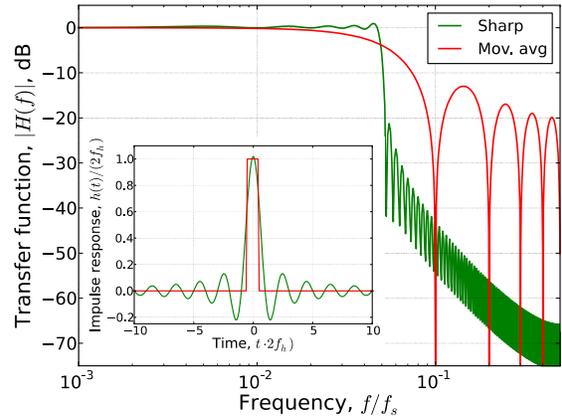

Figure 4: the transfer function plot shows that the selectivity of the truncated sinc filter (in green) is significantly higher than the moving average filter (in red) with the same bandwidth $f_h = 0.05\, f_s$. This is obtained at the expense of the impulse response duration ($10/f_h$ in the example) as shown by the inset. The selectivity is proportional to the duration of the impulse response and can be changed according to the needs of the experiment.

It is advisable to iteratively apply filters with progressively decreasing $f_h$. After each filtering stage, the amount of data can be resized according to the Nyquist theorem; this dramatically reduces the storage and computational requirements. Once the filtering and decimation process is completed, i. e. the optimal $f_h$ has been obtained, the Allan deviation can be computed.

Sampling and storing data according to the Nyquist criterion (at 1 kHz rate or similar) may be unpractical during a real experiment and data filtering and decimation is necessary. This can be done in several manners: on the computer that runs the acquisition or directly by the instrument. Phasemeters usually implement sharp filters; the more common dead-time free phase/frequency counters used in the Lambda-mode [23] implement instead a moving average filter. The latter is equivalent to a first order low pass filter with $f_h = 1/(2t_g)$, being $t_g$ the gate time. This operating mode has the advantage of being readily implemented in the counter without requiring any computational effort of the user and allows an agile visual inspection and data-processing; on the other hand, with such a weak filtering, the unsuppressed link noise may still play a role. A practical example and the comparison of the results with truncated sinc filtering and moving average is described in Section IV. We also stress that the ADEV, as well as further filtering, can be applied to such Lambda-averaged data just as it is done with sharper pre-filters.

## IV. APPLICATION TO AN EXPERIMENTAL CASE

Let us consider a practical case in which the described analysis can be applied. Initially, we will describe how to evaluate the performances of a coherent optical link, then we will consider an hypothetical clocks comparison over this link. For completeness, we will describe both the cases where optical or microwave clocks are compared. We analyze data of the coherent optical link developed by the National Metrology Institute in Italy (INRIM). This link has a total length of 642 km and connects many national research facilities of the Country. Its metrological characterization has been performed by looping the link using a single fiber, to have both ends in the same laboratory. The total link length is then 1284 km. The experimental realization and the link validation are detailed in [3]. Both the electronics apparatus and the measurement instruments are entirely composed of commercial devices and off-the-shelf components. In particular, we realized fully-analog phase-locked loops with clean-up voltage-controlled oscillators; the data are collected using a dead-time free phase/frequency counter [23] in the phasemeter mode at a 1 kHz sampling frequency. The set of data we analyze here lasts 20000 s and contains 20 Msamples (about 1 GB of text data). We removed 33 cycle slips from our dataset, that degraded the performance by about one order of magnitude in the long term; this operation does not affect the results we propose and its only purpose is to increase the length of the cycles-slips free dataset useful for analysis.

The first aspect to consider is the power spectral density of the link residual phase noise, shown in Fig. 4. It contains many cases of interest: for $f > 50$ Hz, the noise is decreasing quickly; for $5$ Hz $< f < 50$ Hz, there is a strong bump (20 dB in this experiment); for $0.5$ Hz $< f < 5$ Hz the noise is dominated by WPN ($\alpha = 0$); for $50$ mHz $< f < 0.5$ Hz the noise is dominated by BPN ($\alpha = +1$); for $1$ mHz $< f < 20$ mHz, there is flicker phase noise ($\alpha = -1$); at lower frequencies, not shown in the figure, we find the signature of environmental effects, such as temperature variations and transients. For the ADEV computation, the most problematic spectral region is between 20 mHz and 50 Hz. To show the effect of each of these typical cases, we apply the proposed method by progressively reducing the bandwidth from 500 Hz to 5 mHz in steps of a factor 10. The results are shown in Fig. 5, together with those obtained with the MDEV. We note that the ADEV is only shown for $\tau \geq 1/(2f_h)$, where it is not affected by the filtering process.

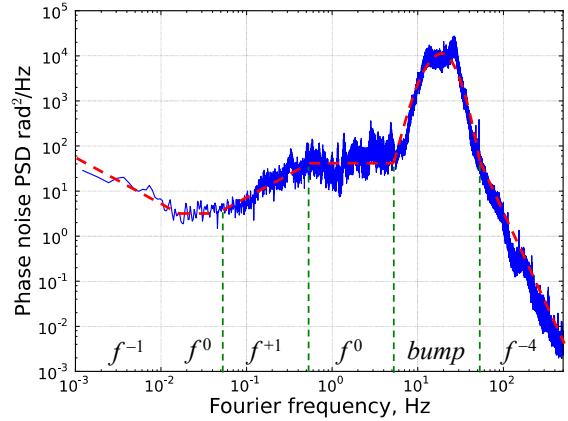

Fig. 5: The phase noise power spectral density (PSD) of the 1284 km optical link. Dashed lines represent the estrapolated noise components.

It can be noticed that the bandwidth reduction from 500 Hz to 50 Hz has no impact on the ADEV estimation, as in this spectral region the spectrum decreases quickly. This stability can also be considered a good approximation of the full-bandwidth case and is described by the relation $\sigma_y(\tau) = 5 \times 10^{-13} \ \tau^{-1}$.

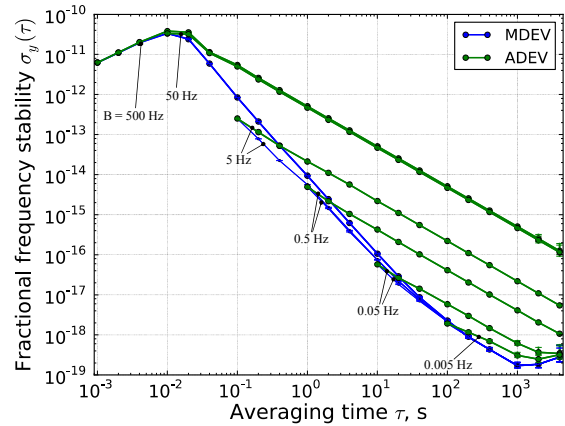

Fig. 6: The ADEV and MDEV (green and blue thin lines respectively) of the 1284 km link with different measurement bandwidths and the unfiltered MDEV (thick blue line). The effect of $f_h$ on MDEV is shown by the thin blue lines.

With filtering, the ADEV estimation improves progressively. A factor 20 improvement is obtained when the bandwidth is reduced from 50 Hz to 5 Hz, i.e. when the bump is filtered out. Further filtering still improves the ADEV depending on the noise type: a factor 5 ($\alpha \cong 0$), then a factor 7.2 ($\alpha \cong +1$) and in the end a factor 2. In the last curve, obtained with $f_h = 5$ mHz, the fast noise has been completely removed, and the ADEV can be considered representative of the link noise. Thus, it is possible to conclude that the link has a resolution of $3 \times 10^{-19}$ at 1000 s with $f_h = 5$ mHz.

We note that the filtered ADEV provides better results than the MDEV for $\tau = 1/f_h$, just because the pre-filter is more selective than the moving average embedded in the MDEV. This is particularly evident for $f_h$ = 5 Hz, where the bump has been filtered out. At 100 ms, it provides an estimation that is 10 dB lower. We note as well that also the MDEV benefits from the filtering process (thin blue lines). It is interesting to take a closer look to the case of $f_h$ = 5 Hz: the unfiltered MDEV takes about two decades (from 100 ms to 10 s) to reach the estimations of the filtered MDEV. In the other cases, the difference is not significant, since the MDEV algorithm well manages noise processes with $\alpha \leq +1$.

## V. CLOCKS COMPARISON AND DISSEMINATION

Once implemented and characterized, the link can be used to transfer stable and accurate frequency references as those generated by atomic clocks. Currently, our clocks ensemble at INRIM is composed of several hydrogen masers, a Cs cryogenic fountain with an accuracy of $1.7 \times 10^{-16}$ and a stability of $2.4 \times 10^{-13} \tau^{-1/2}$ [27], and a Yb lattice clock under development [28]. The primary Cs standard is disseminated through the optical link via an ultrastable laser, whose frequency is referred to the Cs frequency through an optical comb. Evidently, the same can be done using the optical clock as a reference. The main difference in the two cases relies in the frequency stability performances, being at the level of $2.4 \times 10^{-13} \tau^{-1/2}$ for a microwave clock [27], and as low as $1.7 \times 10^{-16} \tau^{-1/2}$ for an optical clock [28-29].

In this section, we consider both the optical and microwave dissemination along an optical link. In particular, each clock is simulated superimposing to the actual link noise a phase random walk obtained by integrating a WFN with Gaussian distribution; for the sake of clarity, here we assume a stability of $1 \times 10^{-13} \tau^{-1/2}$ and $1 \times 10^{-16} \tau^{-1/2}$ for the microwave and optical clocks respectively. Then, the data have been processed by the pre-filtering technique explained in previous sections.

In Fig. 6 we show the phase noise of the signal received by the remote observer. It is evident from the $f^{-2}$ behavior of the spectrum that the information related to the clock arises for frequencies lower than 5 Hz for the microwave clock and 25 mHz for the optical clock. For higher frequencies, the spectrum is dominated by the residual phase noise of the link. It is then natural to choose these frequencies as $f_h$ in the two cases, so that the link contribution is filtered out and the data contain the clock information only.

Then, we computed the ADEV, the MDEV and the theoretical deviations for the two clocks. The results are shown in Fig. 7. In the case of the optical clock, the remote user can recover the clock stability after 20 s of averaging time without bias.

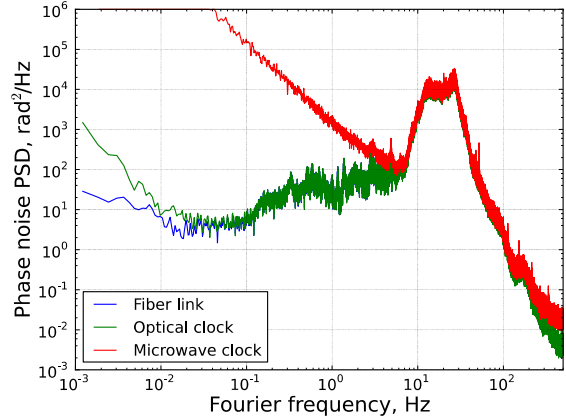

Fig. 7: the phase noise power spectral density of the bare link (blue line) and of a link that carries the microwave and optical clock signals in the remote laboratory (red and green line respectively).

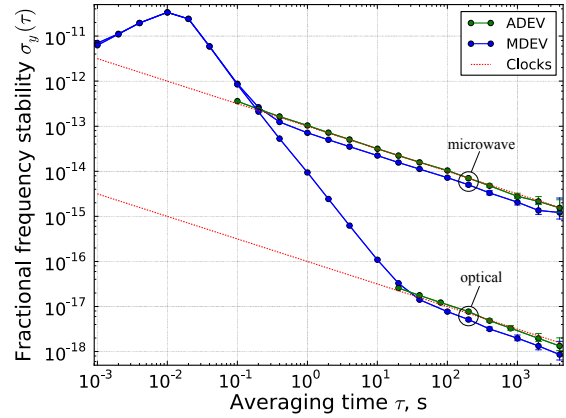

Fig. 8: the stability of the clock signals in the remote laboratory when either the MDEV (blue lines) or the ADEV associated with filtering (green lines) are used. The uppermost curves represent the case of the microwave clock; in this case the ADEV is computed with $f_h$ = 5 Hz. The lower curves are related to the optical clock; in this case the ADEV is computed with $f_h$ = 25 mHz. The dotted red lines represent the stability of the two clocks.

On the other hand, the MDEV provides a −3 dB biased estimate of the clock stability and in addition it is representative of the clock stability only after 40 s. The advantage in the case of the microwave clock is more evident. Here the filtered ADEV represents the clock stability at the remote site from 100 ms, while the full bandwidth MDEV carries the information of the clock stability from an averaging time of 400 ms. We point out that, at 100 ms, the ADEV well estimates the clock stability, while the MDEV estimation is 7.5 dB higher. This is due to the fact that, in this case, $f_h$ is close to the bump, which is effectively rejected by the pre-filtering process.

As a last consideration, we show in more detail the benefit of sharp filtering on the link noise. The filtering process used to estimate the link performances is the result of cascaded application of a truncated sinc filter with $f_h = 0.05\ f_s$. After each filtering stage, data have been decimated by a factor 10. Fig. 9 shows the link phase noise when filtered with the sinc filter and with a moving average. From the picture, it can be seen that the moving average is not selective enough to completely reject the link noise for $f > f_h$, thus affecting the ADEV computation.

In addition, care must be taken in the decimation process of simply averaged data, as it could lead to a peculiar type of aliasing. Typically, one expects aliasing to degrade the white noise level; if so, it would completely bury the link noise at low frequency. However, thanks to the first order zeroes of the moving average transfer function (at $f = n/t_g$, with n = 1, 2,…), the aliasing contributes as $f^2$ for $f \ll 1/t_g$, therefore becoming negligible with respect to the link noise. A comparison of the ADEV obtained with sinc filtering and moving average is shown in Fig. 10. A higher stability is noticed when the moving average is used: this is expected, given the lower selectivity of this filter, nevertheless, at 5 mHz, far from the bump, the results obtained are very close to the ones calculated by using the sharp filter.

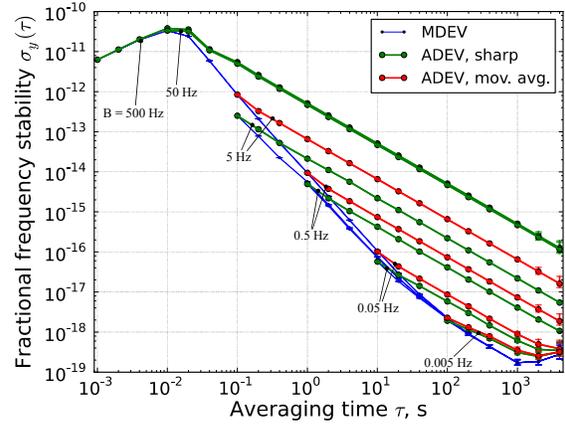

Figure 10: the ADEV computed from sinc-filtered data (green curves) or moving-average-filtered data (red curves) for different bandwidths.

## V. Conclusion

In this paper, we have shown how the Allan deviation can be used as a statistical estimator of the optical links performances. In particular, we focused on the fact that optical links are mainly intended to compare atomic clocks or to disseminate frequency. Our approach makes use of a preliminary data filtering, to reject the noise of the optical fiber link. The ADEV is then applied to filtered data, to get an unbiased estimation of uncertainty. The pre-filtering approach is very powerful as it enables to achieve the desired resolution with the ADEV already at short averaging times. The filter bandwidth and selectivity can be tailored on the specific experiment, depending on the link noise and on the clocks to be compared. On the contrary, the MDEV still suffers from the residual link noise at these timescales.

We applied this technique to analyze experimental data of the 1284 km optical fiber link for frequency dissemination developed by the National Institute of Metrology in Italy. We evaluated the link performance and we simulated two clocks comparisons over this fiber. We have shown how pre-filtering associated with the ADEV is a performing method to characterize the links stability.

This approach can be useful in view of an increasing number of clocks comparisons via optical fiber, also considering that it is already implemented in commercial devices such as dead-time free phase/frequency counters operated in the Lambda mode.

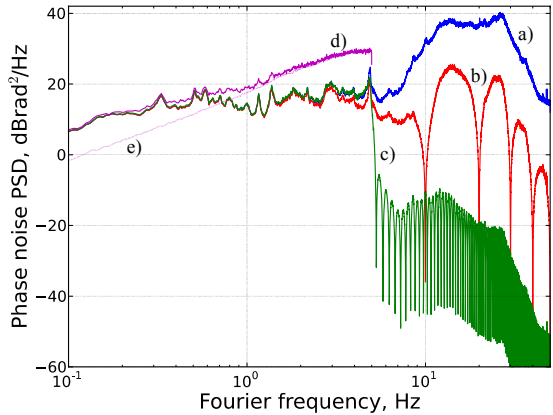

Figure 9: the unfiltered (blue line, a) and filtered link noise when either a moving average(red line, b) or a sinc filter (green line, c) are applied, both with the same $f_h$ = 5 Hz. The thin purple line (d) represents the PSD of data corresponding to the red curve after decimation. The effect of aliasing is clearly visible between 1 Hz and 5 Hz. The dotted line, e) reports the aliasing component and demonstrates its slope.

### Acknowledgments

The authors thank the Optical Link team at INRIM for useful discussion and for allowing use of the experimental data and Marco Pizzocaro for useful suggestions and careful reading of the Manuscript.

This work was supported by the Italian Ministry of Research and Education under the LIFT project of the Progetti Premiali program and by the European Metrology Research Program (EMRP project SIB55-ITOC). The EMRP is jointly funded by the EMRP participating countries within EURAMET and the European Union.

REFERENCES

[1] O. Lopez, A. Haboucha, B. Chanteau, C. Chardonnet, A. Amy-Klein, G. Santarelli, "Ultra-stable long distance optical frequency distribution using the Internet fiber network," Opt. Express Vol. 20, pp. 23518-23526, Sept. 2012.

[2] S. Droste, F. Ozimek, Th Udem, K. Predehl, T.W. Hansch, H. Schnatz, G. Grosche, R. Holzwarth, "Optical-Frequency Transfer over a Single-Span 1840 km Fiber Link," Phys. Rev. Lett. Vol. 111, 110801, Sept. 2013.

[3] D. Calonico, E. K. Bertacco, C. E. Calosso, C. Clivati, G. A. Costanzo, M. Frittelli, A. Godone, A. Mura, N. Poli, D. V. Sutyrin, G. Tino, M. E. Zucco, F. Levi "High-accuracy coherent optical frequency transfer over a doubled 642-km fiber link," Appl. Phys. B Vol. 117, pp. 979-986, Aug. 2014.

[4] K. Predehl, G. Grosche, S. M. F. Raupach, S. Droste, O. Terra, J. Alnis, Th. Legero,T. W. Hänsch, Th. Udem, R. Holzwarth, H. Schnatz, "A 920-Kilometer optical fiber link for Frequency Metrology at the 19th Decimal Place," Science Vol. 336, pp. 441-444, Apr. 2012.

[5] M. Fujieda, M. Kumagai, S. Nagano, A. Yamaguchi, H. Hachisu, and T. Ido, "All optical link for direct comparison of distant optical clocks," Opt. Express Vol. 19, pp. 16498-16507, Aug. 2011.

[6] A. Yamaguchi, M. Fujieda, M. Kumagai, H. Hachisu, S. Nagano, Y. Li, T. Ido, T. Takano, M. Takamoto, and H. Katori, "Direct comparison of distant optical lattice clocks at the $10^{-16}$ uncertainty," Appl. Phys. Express 4, 082203, Aug. 2011.

[7] N. Chiodo, N. Quintin, F. Stefani, F. Wiotte, C. Salomon, A. Amy-Klein et al. "4-Span Cascaded Optical Link of 1500 km Using the Internet Fiber Network," proc. of the Joint IFCS and EFTF, Denver, April 14-17, 2015.

[8] S. M. F. Raupach, A. Koczwara, and G. Grosche, Brillouin amplification supports $1\times10^{-20}$ accuracy in optical frequency transfer over 1400 km of underground fibre, arXiv:1504.01567, 2015.

[9] D. Calonico, M. Inguscio and F. Levi, "Light and the distribution of time," Europ. Phys. Journ., in press.

[10] A. Matveev, C. G. Parthey, K. Predehl, J. Alnis, A. Beyer, R. Holzwarth, Th. Udem, T. Wilken, N. Kolachevsky, M. Abgrall, D. Rovera, C. Salomon, P. Laurent, G. Grosche, O. Terra, T. Legero, H. Schnatz, S. Weyers, B. Altschul, T. W. Hansch, "Precision Measurement of the Hydrogen 1S -2S Frequency via a 920-km Fiber Link" Phys. Rev. Lett. 110, 230801, June 2013.

[11] C. Clivati, D. Calonico, M. Frittelli, A. Mura, F. Levi, "An optical fiber link for the remote comparison of optical clocks and geodesy experiments," proc. of the Joint International Frequency and Control Symposium and European Frequency Control Forum, Denver, 12-16 April 2015

[12] C. Clivati, R. Ambrosini, C. Bortolotti, G. A. Costanzo, M. Frittelli, F. Levi, A. Mura, F. Perini, M. Roma, M. E. Zucco and D. Calonico, "The Optical Fiber Link LIFT for Radioastronomy," proc. of the Joint IFCS and EFTF, Denver, April 14-17, 2015.

[13] L.-S. Ma, P. Jungner, J. Ye, J.L. Hall, "Delivering the same optical frequency at two places: accurate cancellation of phase noise introduced by an optical fiber or other time-varying path," Opt. Lett. Vol. 19, pp. 1777-1779, Nov. 1994.

[14] C. E. Calosso, E. Bertacco, D. Calonico, C. Clivati, G. A. Costanzo, M. Frittelli, F. Levi, A. Mura, and A. Godone, "Frequency transfer via a two-way optical phase comparison on a multiplexed fiber network," Opt. Lett. vol. 39, pp. 1177-1180, March 2014.

[15] P. A. Williams, W. C. Swann, and N. R. Newbury, "High-stability transfer of an optical frequency over long fiber-optic links," J. Opt. Soc. Am. B Vol. 25, pp. 1284-1293, Aug. 2008.

[16] D. W. Allan, "Statistics of Atomic Frequency Standards," Proc. IEEE Vol. 54, pp. 221-230, 1966.

[17] D. W. Allan, "Should the Classical Variance Be Used As a Basic Measure in Standards Metrology?" IEEE Trans. Instrum. Meas. IM-36, pp. 646-656, June 1987.

[18] D. W. Allan and J. Barnes, "A modified "Allan variance" with increased oscillator characterization ability," in Proceedings of the 35th Ann. Freq. Control Symposium, pp. 470–475.

[19] E. Rubiola "Phase noise in Oscillators" 1st ed. (Cambridge University Press, 2009).

[20] E. Benkler, C. Lisdat, U. Sterr "On the relation between uncertainties of weighted frequency averages and the various types of Allan deviations," Arxive 1504.00466, 2015.

[21] C. E. Calosso, "Digital Phasemeter in time and frequency metrology," proc. of the Joint UFFC, EFTF and PFM Symposium (IEEE, 2013), p. 747–749.

[22] C. E. Calosso, E. K. Bertacco, D. Calonico, C. Clivati, G. A. Costanzo, M. Frittelli, F. Levi, S. Micalizio, A. Mura, and A. Godone, "Doppler-stabilized fiber link with 6 dB noise improvement beyond the classical limit," Opt. Lett. Vol. 40, pp. 131-134, Jan 2015.

[23] G. Kramer, W. Klische, "Extra High precision digital phase recorder," in Proceedings of the 2001 IEEE International Frequency Control Symposium and PDA Exhibition, Seattle, WA, 2001, p. 144

[24] Mark Owen (2007) "Practical signal processing" Cambridge University Press. p. 81. ISBN 978-0-521-85478-8.

[25] E. Rubiola, "On the measurement of frequency and of its sample variance with high-resolution counters," Rev. Sci. Instrum. vol. 76, p. 054703-6, 2005.

[26] S. T. Dawkins, J. J. McFerran, A. N. Luiten "Considerations on the Measurement of the Stability of Oscillators with Frequency Counters," IEEE Trans. Ultrason. Ferroelectr. Freq. Control, vol. 54, pp. 918-925, 2007.

[27] F. Levi, D. Calonico, C. E. Calosso, A. Godone, S. Micalizio and G. A. Costanzo, "Accuracy Evaluation of ITCsF2: a nitrogen-cooled caesium fountain," Metrologia vol. 51, p. 270 (2014).

[28] M. Pizzocaro, in Proceedings of the European Frequency and Time Forum, Prague, 21-25 July, 2013.

[29] N. Hinkley, J. A. Sherman, N. B. Phillips, M. Schioppo, N. D. Lemke, K. Beloy, M. Pizzocaro, C. W. Oates, A. D. Ludlow, "An Atomic Clock with $10^{-18}$ Instability," Science vol. 341, pp. 1215-1218 (2013).